\documentclass[preprint,12pt]{elsarticle}

\usepackage{lineno}
\modulolinenumbers[5]

\usepackage{hyperref}

\journal{Computer Physics Communications}
\newcounter{bla}

\usepackage{amssymb}
\usepackage{graphicx}
\usepackage{amsfonts}
\usepackage{amsmath}
\usepackage{soul}
\usepackage{xcolor}

\begin{document}

\begin{frontmatter}

\title{Parallelized Discrete Exterior Calculus for Three-Dimensional Elliptic Problems}
\author{Pieter D. Boom}\corref{boom}
\author{Ashley Seepujak}
\author{Odysseas Kosmas}
\author{\ \\Lee Margetts}
\author{Andrey Jivkov}
% \ead{}
\cortext[boom]{Corresponding author.\\\textit{E-mail address:} pieter.boom@manchester.ac.uk}
\address{Department of Mechanical, Aerospace and Civil Engineering, The University of Manchester, George Begg Building, M1 3BB Manchester, UK.}

\begin{abstract}

A formulation of elliptic boundary value problems is used to develop the first discrete exterior calculus (DEC) library for massively parallel computations with 3D domains. This can be used for steady-state analysis of any physical process driven by the gradient of a scalar quantity, e.g. temperature, concentration, pressure or electric potential, and is easily extendable to transient analysis. In addition to offering this library to the community, we demonstrate one important benefit from the DEC formulation: effortless introduction of strong heterogeneities and discontinuities. These are typical for real materials, but challenging for widely used domain discretization schemes, such as finite elements. Specifically, we demonstrate the efficiency of the method for calculating the evolution of thermal conductivity of a solid with a growing crack population. Future development of the library will deal with transient problems, and more importantly with processes driven by gradients of vector quantities. 
\end{abstract}

%%%%%%%%%%%%%%%%%%%%%%%%%%%%%%%%%%%%%%%%%%%%%%%%%%%%%%%%%%%%%%%%%%%%%%%%%%%%%%%%%%%%%%%%%%%
%%%%%%%%%%%%%%%%%%%%%%%%%%%%%%%%%%%%%%%%%%%%%%%%%%%%%%%%%%%%%%%%%%%%%%%%%%%%%%%%%%%%%%%%%%%
\end{frontmatter}

%\linenumbers

\noindent
{\bf PROGRAM SUMMARY}\\[1ex]
\begin{small}
\noindent
{\em Program Title:} ParaGEMS                                         \\[1ex]
{\em CPC Library link to program files:} (to be added by Technical Editor) \\[1ex]
{\em Developer's repository link:} https://bitbucket.org/pieterboom/paragems/ \\[1ex]
{\em Code Ocean capsule:} (to be added by Technical Editor)\\[1ex]
{\em Licensing provisions:} BSD 2-clause  \\[1ex]
{\em Programming language:} Fortran 90                \\[1ex]
{\em External routines/libraries:} BLAS/LAPACK, MPI, PETSc [1], hypre ParaSails [2,3] \\[1ex]
{\em Nature of problem:} Large scale simulation of 3D elliptic boundary value problems in discrete media with heterogeneous or discontinuous properties. \\[1ex]
  %Describe the nature of the problem here. \\
{\em Solution method:} ParaGEMS is a library implementing discrete exterior calculus (DEC) for distributed memory architectures. It is distributed with mini-applications for solving 3D elliptic boundary value problems, which describe a number of physical laws (Fourier's law, Fick's law, Darcy's law and Ohm's law), in discrete media with heterogeneous or discontinuous properties. \\[1ex]
  %Describe the method solution here.
{\em Additional comments including restrictions and unusual features:} ParaGEMS requires meshes generated in the default format used by Triangle [4] and TetGen [5].\\

\end{small}

%---------------------------------------------------------------------------------------------------------------------------------------
\section{Introduction}
\label{Sec_Introduction}
%---------------------------------------------------------------------------------------------------------------------------------------

A number of physical laws for continuous media are described by mathematically identical equations: Fourier's law for heat conduction, Fick's law for mass diffusion, Darcy's law for fluid flow through porous media and Ohm's law for electricity. In these, the flux density of a given quantity (heat, amount of substance, fluid volume, electric current) is proportional to the gradient of a scalar quantity (temperature, concentration, pressure, electric potential). The coefficient of proportionality is a material property (thermal conductivity, diffusivity, permeability, electric conductivity), which may be constant across the domain of interest for homogeneous domains, or spatially-varying for heterogeneous domains. The conservation of the scalar quantity is ensured by equating its temporal derivative to the divergence of the corresponding flux density. This leads to parabolic partial differential equations (PDE) for the processes within the domain. Initial boundary value problems are then formulated by prescribing initial conditions for the scalar quantity as well as boundary conditions for the scalar quantity (Dirichlet) and for the flux density (Neumann). Of particular interest in many cases is the steady-state solution of the problem, for which the equation reduces to an elliptic PDE and the formulation to a boundary value problem (BVP). We will use the heat transfer problem as a proxy for all of these mathematically identical problems. 

Analytical solutions of BVP can be found only for a very limited set of domain geometries, spatial variations of material properties and boundary conditions. These are typically of no practical relevance, but serve as benchmarks for testing a multitude of numerical methods developed for solving BVP with arbitrary geometries, material properties and boundary conditions. One class of methods are based on the discretization of the domain, such as finite elements \cite{Lewis_FEM_1996}, or of its boundary, such as boundary elements \cite{Ibanez_BEM_2002}, and various approximations of the scalar and flux fields. The main feature of this class of methods is that they rely on continuity of the matter in the domain, and represent the parabolic PDE with algebraic equations arising from the approximation of the fields. Real materials may contain strong heterogeneities, such as jumps in properties across internal walls, and discontinuities preventing flux, such as cracks. Whilst modeling of such situations is challenging with methods relying on continuity, it is possible with a substantial amount of additional work to adjust the discretization to the heterogeneities and/or discontinuities present in a particular material. Examples include applying the finite element method \cite{Lu_JMPS_1995, Pestchanyi_FED_2010, Shen_CBM_2015, Shen_CBM_2017} and the boundary element method \cite{Simionato_ZAMM_2019} to problems of heat transfer in heterogeneous and fractured media.

Another class of methods with the potential for a more natural representation of strong heterogeneities and discontinuities, are based upon a discrete representation of matter within the domain. These include discrete element methods \cite{Oschmann_JPE_2016, Peng_PECS_2020}, smoothed particle hydrodynamics \cite{Clearly_JCP_1999, Alshaer_CMS_2017}, and peridynamics \cite{Bobaru_JCP_2012, Yan_JH_2021}. Discrete methods are appropriate for the task at hand, in particular for the link between the micro-properties associated with interactions between the discrete entities and the experimentally-measured macroscopic properties.

One approach to avoid such approximations, yet to allow for a natural representation of heterogeneities and discontinuities to exist or emerge and to grow, is to use a discrete topological representation of the material and discretize the operators describing the processes (gradients) and conservation (divergences). Such an approach, referred to as discrete exterior calculus (DEC) \cite{Hirani_DEC_2003}, draws to some extent upon the smooth exterior calculus of differential forms, but is adapted for discrete topologies. The principal advantage of DEC is that it mimics the coordinate invariance of exterior calculus on smooth manifolds, clearly differentiating between topological, geometric and physical contributions to the discrete operators. The discrete topology is a cell complex in the language of algebraic topology. DEC leads to discrete algebraic equations which obey exactly the topological structure, but the solution is an approximation to a continuum case. In contrast to domain discretization schemes, which use approximations of scalar quantities and flux densities, the approximation in DEC arises from the number of cells in the analyzed domain. Considering that the continuum may not be a sufficiently good representation of a real material, and that a particular cell complex may represent sufficiently well the internal structure of this material \cite{Belayachi_JBE_2019}, it could be argued whether the term approximation is indeed appropriate in all cases. Nevertheless, it has been shown recently that DEC for scalar problems, such as those considered herein, converges to the continuum solution with decreasing cell size \cite{Schulz_DCG_2020}.

It should be noted, that smooth exterior calculus has been used for a separate development, referred to as finite element exterior calculus \cite{Arnold_CBMS_2018}. This is an effective structure-preserving technique which leads to canonical constructions of stable finite elements, thus improving significantly the technology of the finite element method. Whilst the method has been applied to heat conduction in cracked media \cite{Vu_IJHMT_2015}, the method inherits the challenges of the classical finite element\st{s} approach when dealing with strong heterogeneities and discontinuities, which can be overcome by operating on discrete topological spaces as in DEC.

The use of DEC for scalar problems, particularly for Darcy's flow in 2D and 3D domains has been previously reported \cite{Hirani_IJCMSM_2015}. The present work builds upon this development to produce an efficient parallelized solver for all 3D elliptic problems. The software implementation is capable of running on distributed memory high-performance computing (HPC) architectures to facilitate solving large-scale problems. The open-source software TetGen \cite{Si_ACM_2015} is used to generate simplicial complexes from which their Voronoi duals are constructed to represent the microstructures of 3D solids. The topology and geometry of these complexes are used to construct discrete versions of the smooth differential operators involved in the continuum formulation of BVP. Material properties are introduced as a discrete field over the volumes or faces of the complexes, allowing in principle the effortless introduction of heterogeneities and discontinuities at any stage of a simulation. 

A specific example is selected to demonstrate the new software: heat conduction in a domain of cubical shape. Analyses of scalability with respect to the number of cores used for simulation, and of stability with respect to the domain discretization are presented. Models are then used to study the effect of a growing crack population on the effective thermal conductivity of the material domain. Two different criteria for cracking (face conductivity change from finite value to zero) are tested - the emergence of spatially random cracks, and the emergence of cracks at faces with maximal heat flux density. It should be noted, that the problem of thermal conduction in cracked solids has been a subject of analytical studies with associated constraints on domain and crack geometries and spatial distributions: an infinitely large plate with a single crack \cite{Kit_JEP_1972, Glushko_AA_2016}, a plate with isotropic distribution of identical cracks \cite{Hoenig_JCM_1983}, cracks with  specific geometries \cite{Savruk_MS_1987, Hasselman_JCM_1987, Benveniste_JAP_1989}, and cracks in domains with specific geometry \cite{Hu_IJHTMT_2013, Tran_JAG_2018, Yan_IJNAMG_2019}. The proposed formulation avoids all constraints and allows for analysis of heterogeneous, anisotropic materials with an arbitrary criterion for the evolution of properties.

The paper is organized as follows. Section \ref{Sec_Methodology} describes the parallel implementation of DEC in the software library ParaGEMS, along with a DEC formulation of heat conduction and a method of introducing cracks into the material. Numerical simulations are presented in Section \ref{Sec_Results_and_Discussion} to verify the implementation and to evaluate parallel performance. It also presents simulations using both deterministic and stochastic cracking processes. Finally, in Section \ref{Sec_Conclusions_and_perspectives}, conclusions and perspectives are drawn, highlighting the directions for future improvements and developments.

%---------------------------------------------------------------------------------------------------------------------------------------
\section{Methodology}
\label{Sec_Methodology}
%--------------------------------------------------------------------------------------------------------------------------------------- 
This section presents a new software library implementing discrete exterior calculus for distributed memory architectures called ParaGEMS, and its application to scalar problems with material cracking. The parallelization strategy aims to keep computations as local as possible, at the expense of storing some duplicate information at interfaces between cells on adjacent processes. This can be mitigated to some extent by careful grouping and allocation of cells to processes, which also minimizes parallel communication. Local geometric computations in ParaGEMS are inspired by procedures in the serial software library PyDEC developed by Bell and Hirani \cite{Bell_PYDEC_2012}. The DEC formulation for Darcy flow presented by Hirani {\it et al.} \cite{Hirani_IJCMSM_2015} is extended to include cracking media with minimal modification of the system matrix for each new crack introduced. Finally, parallel solution of the resulting linear systems is obtained using PETSc \cite{petsc-web-page}. 

%---------------------------------------------------------------------------------------------------------------------------------------
\subsection{Problem description}
\label{subSEC_Physical_problem}
%--------------------------------------------------------------------------------------------------------------------------------------- 

We consider an open domain $M \in \mathbb{R}^{3}$, with closure $\bar{M}$ and boundary $\partial M:=\bar{M} \setminus M$. With the classical vector calculus notations, the heat flux density in $M$ is given by the Fourier law 
\begin{equation} \label{Fourier}
    \mathbf{f} \;\; = \;\; -\kappa \;\; \nabla T,
\end{equation}
\noindent where $T$ is the temperature, and $\kappa$ is the thermal conductivity at any given point in $M$, and the conservation of energy for steady-state heat conduction in $M$, is 
\begin{equation} \label{Conservation}
    \nabla \; \cdot \; \mathbf{f} \;\; = \;\; 0.
\end{equation}

A boundary value problem is formulated by supplementing conditions on $\partial M$. The boundary is represented as a union of non-overlapping sub-domains, $\partial M_D$ and $\partial M_N$, i.e. $\partial M_D \cup \partial M_N = \partial M$ and $\partial M_D \cap \partial M_N = \varnothing$. Dirichlet and Neumann boundary conditions are prescribed, respectively, on the two sub-domains by
\begin{eqnarray}
\label{Dirichlet}
    T &=& T_0 \quad \quad \textrm{on} \quad \partial M_D,\\
\label{Neumann}
    \mathbf{f} \cdot \mathbf{n} &=& f_0 \quad \quad \textrm{on} \quad \partial M_N,
\end{eqnarray}
\noindent where $T_0$ is prescribed temperature at a given point on $\partial M_D$, $\mathbf{n}$ is the unit normal to a given point on $\partial M_N$, and $f_0$ is prescribed magnitude of heat flux density normal to a given point on $\partial M_N$.

The exterior calculus formulation uses the notion of primal and dual spaces for differentiating between domains of primal variables and flux densities, respectively. Furthermore, it uses musical isomorphisms to translate vector fields into differential forms (the flat operator $\flat$) and vice versa (the sharp operator $\sharp$). For the BVP defined above, the temperature is a 0-form, $\alpha$, on the primal space, and the flux density is a 2-form $\omega$ on the dual space, which is a flat of the vector field $\mathbf{f}$, i.e. $\omega:=\mathbf{f}^{\flat}$. The exterior derivatives of differential forms on both spaces, denoted by $d$, are defined in a canonical way from the topology, while the connections between the primal and dual spaces are given by Hodge star operators, denoted by $\star$, that contain geometric and physical properties of the material. The specific translation of the BVP into the language of exterior calculus of differential forms reads 
\begin{eqnarray}
\label{Fourier_EXT}
    \omega &=& \star d \alpha \quad \quad \textrm{in} \quad M,\\
\label{Conservation_EXT}
    \star d \omega &=& 0 \qquad \quad \textrm{in} \quad M,\\
\label{Dirichlet_EXT}
    \alpha &=& T_0 \qquad \quad \textrm{on} \quad \partial M_D,\\
\label{Neumann_EXT}
    \omega &=& f_0 \qquad \quad \textrm{on} \quad \partial M_N,
\end{eqnarray}
\noindent where the thermal conductivity $\kappa$ is incorporated into the Hodge-star operator of Eqn. \ref{Fourier_EXT}, hence it is associate with the pair of a primal 1-form $d \alpha$ and its dual 2-form $\omega$.

DEC uses this exterior calculus formulation of the BVP by specific translation of the exterior derivatives and Hodge-star operators for discrete topological spaces, referred to as cell complexes. Specifically, we use simplicial and polyhedral cell complexes, which are dual to each other. In the language of algebraic topology, our cell complexes are 3-complexes, containing 0-cells (nodes), 1-cells (edges), 2-cells (faces), and 3-cells (volumes). Let $\Omega$ denote one such 3-complex. Linear combinations of $p$-cells in $\Omega$ are called $p$-chains and play the role of integration domains. Functions on $p$-cells in $\Omega$ are called $p$-cochains and play the role of integrands. The space of all $p$-chains is a vector space denoted by $C_{p}\left( \Omega \right)$; the space of all $p$-cochains is a vector space denoted by $C^{p}\left( \Omega \right)$. The topology of $\Omega$, i.e. the connectivity of different cells, together with orientations of these cells, define operators for mapping $p$-chains to $(p-1)$-chains, referred to as the boundary operator $\partial$, and $p$-cochains to $(p+1)$-cochains, referred to as the coboundary operator $\delta$. DEC assumes that the $p$-cochains are discrete analogues of differential forms and the coboundary operator is discrete analogue of the exterior derivative in the smooth exterior calculus, i.e. $d \equiv \delta$. This assumption appears to be sufficient for scalar problems, such as those considered in the present work. The boundary and coboundary operators give rise to the following chain and cochain complexes 
\begin{equation}
0 \xrightarrow{} C_3\left(\Omega\right) \xrightarrow{\partial} C_2\left(\Omega\right) \xrightarrow{\partial} C_1\left(\Omega\right) \xrightarrow{\partial} C_{0}\left(\Omega\right)\xrightarrow{} 0,
\end{equation}
\begin{equation}
0 \xrightarrow{} C^0\left(\Omega\right) \xrightarrow{d} C^1\left(\Omega\right) \xrightarrow{d} C^2\left(\Omega\right) \xrightarrow{d} C^3\left(\Omega\right)\xrightarrow{} 0,
\end{equation}
\noindent where $\partial \circ \partial = 0$ and $d \circ d = 0$, representing the topological property that the boundary (coboundary) of a boundary (coboundary) is empty. Importantly, the dual of a given $\Omega$ has inverted operators, i.e. the boundary operator on $\Omega$ is a coboundary operator on the dual, and the coboundary operators on $\Omega$ is a boundary operator on the dual. The geometrical features of the 3-complex are incorporated into the discrete Hodge-star operators. Considering $\Omega$ to be a simplicial 3-complex (Delauney triangulation of a domain) and constructing its circumcentic dual (Voronoi dual) leads to simple expressions for the Hodge-star operators as diagonal matrices. These are formed by ratios of the volumes of corresponding dual and primal cells. Details can be found in \cite{Hirani_DEC_2003}.

The use of these structures for the problem at hand is as follows. The temperature 0-form, $\alpha$, is a function on the nodes, i.e. a 0-cochain on one of the complexes, its exterior derivative, $d\alpha$, is a function on the edges, or a 1-cochain on the same complex, while the flux density 2-form, $\omega$, is a function on the faces, or a 2-cochain on the dual complex. The thermal conductivity can be considered as a property associated with pairs of primal 1-cell and dual 2-cell, or simply with dual 2-cells. The discrete version of the BVP is a system of algebraic equations, with straightforward implementation of the boundary conditions and introduction of heterogeneities and discontinuities by changing thermal conductivities of dual 2-cells.

%---------------------------------------------------------------------------------------------------------------------------------------
\subsection{Parallelization Strategy}
\label{subSEC_Parallelisation_Strategy}
%--------------------------------------------------------------------------------------------------------------------------------------- 

\begin{figure}[!ht]
    \centering
    \includegraphics[width=0.495\textwidth]{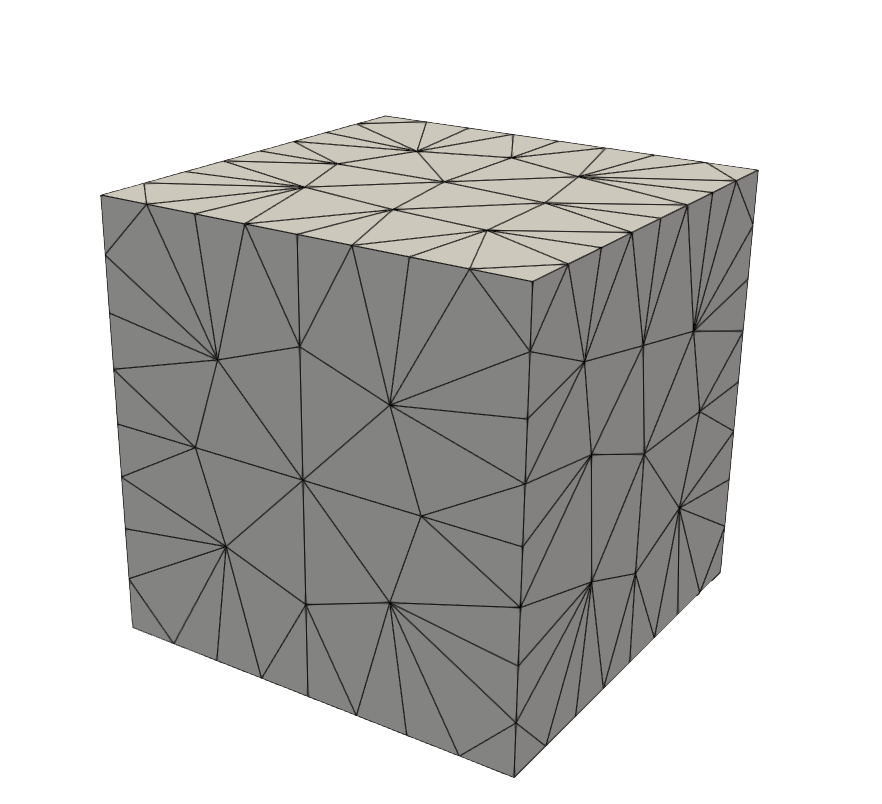}
    \includegraphics[width=0.495\textwidth]{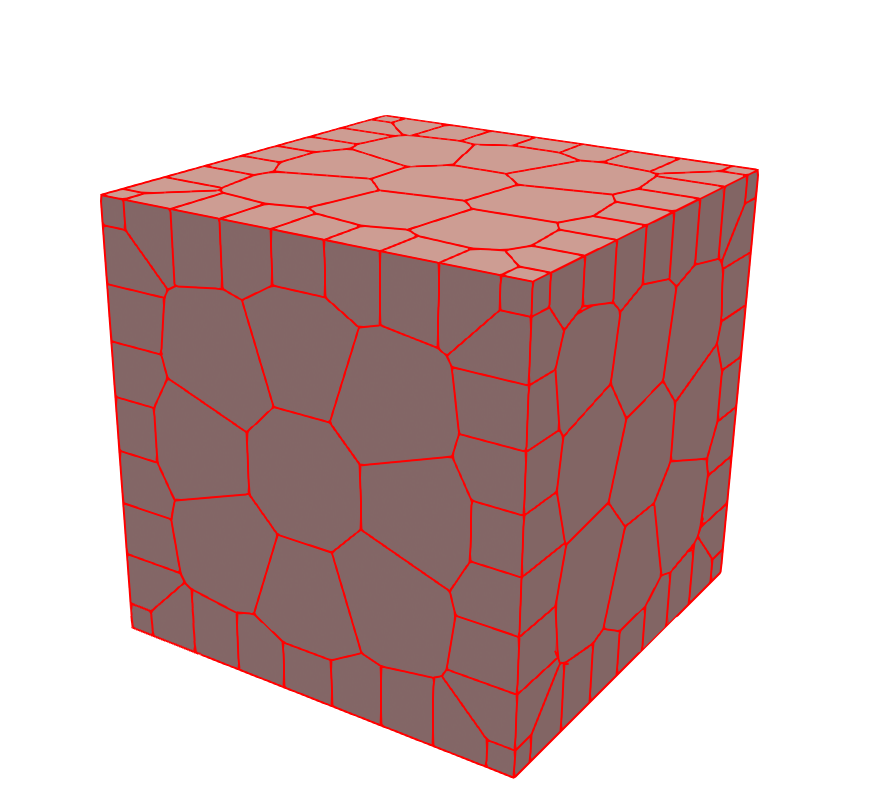}
    \caption{An example Delaunay tetrahedralisation and the corresponding Voronoi diagram generated using TetGen, convert to Wavefront obj format and visualized using ParaView.}
    \label{fig:prml_dual}
\end{figure}

The parallelization strategy in ParaGEMS is based on mesh partitioning. DEC operates on a combination of Delaunay tetrahedral and Voronoi meshes (See Fig. \ref{fig:prml_dual}), which are dual to each other. In this work, the primal is obtained using TetGen \cite{Si_ACM_2015} and the dual constructed as a preprocessing step in ParaGEMS. The use of either the Delaunay or the Voronoi cells to describe the underlying physical phenomena presents slightly different challenges for implementing boundary conditions, but is otherwise largely a matter of choice. With a long-term view towards geometric descriptions of solid mechanics, with cells representing mesoscale grain structures for example, the use of the more geometrically-flexible Voronoi cells is preferred. Therefore, ParaGEMS currently partitions the cells of the Voronoi mesh across the available processes, or equivalently the vertices of the Delaunay mesh. 

\begin{figure}[!ht]
    \centering
    \includegraphics[width=0.495\textwidth]{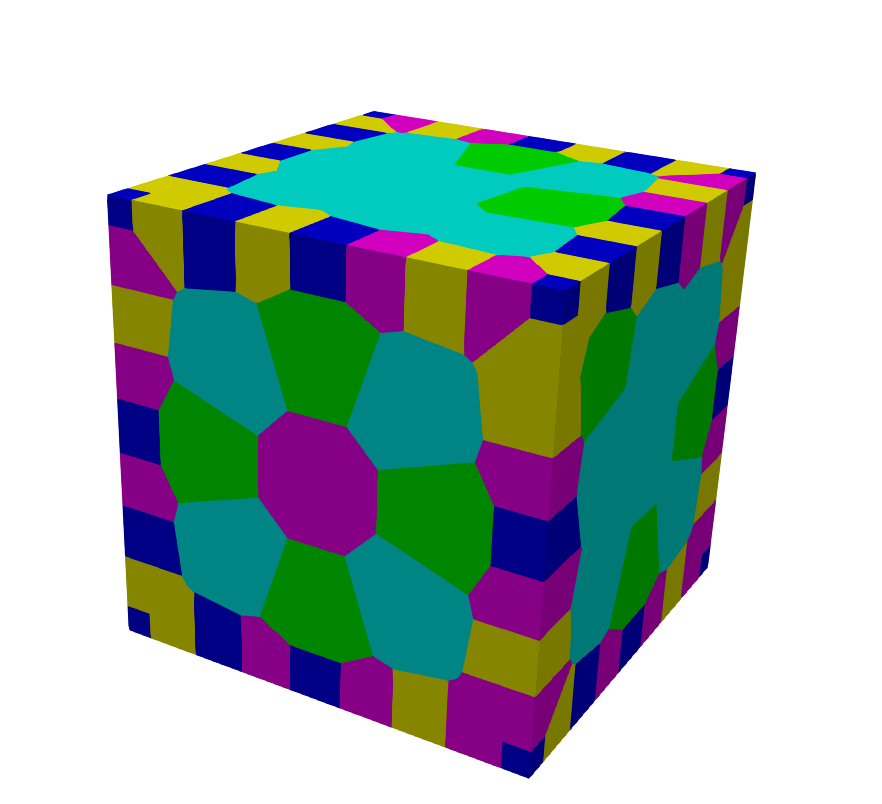}
    \includegraphics[width=0.495\textwidth]{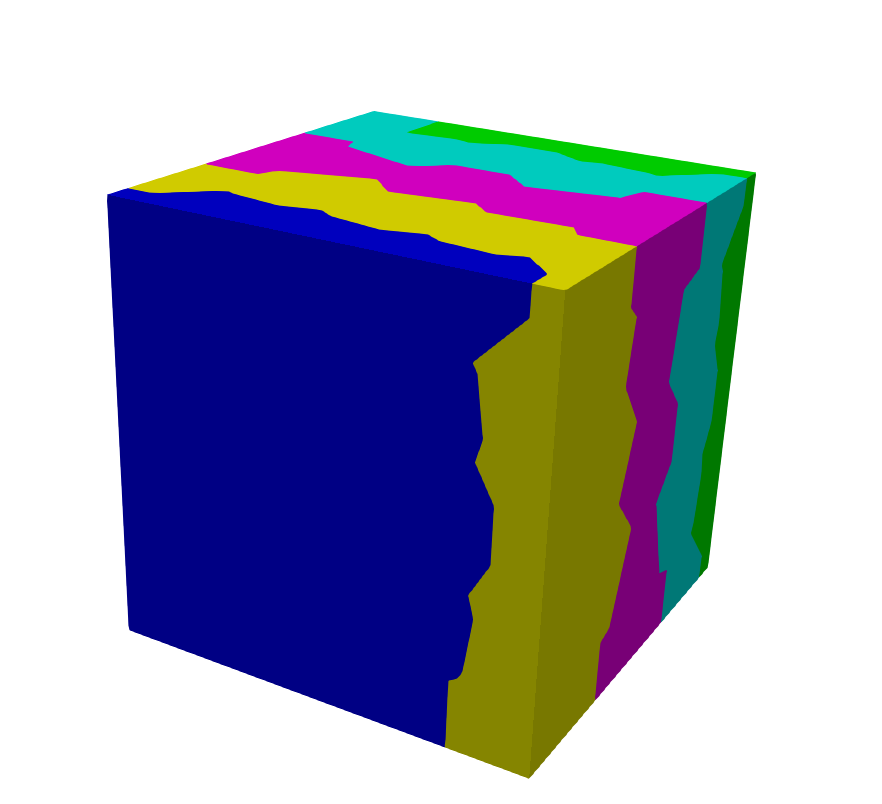}
    \caption{Parallel mesh partitioning using default Voronoi cell ordering from Tetgen (left) and after using the pre-processing tool (right).}
    \label{fig:load_bal}
\end{figure}

Partitioning is done somewhat naively, by dividing the total number of Voronoi cells equally between the different available processes. When non-integer division of cells and processes occurs, a single additional cell is allocated to processes with the lowest rank, beginning with the root process, until the total number is reached. Allocation of cells is done in sequential blocks based upon the numbering provided within the TetGen mesh files. An example is shown in Table \ref{tab:partition}. A pre-processing utility is provided with ParaGEMS which sorts the files, resulting in the Voronoi cells (Delaunay nodes) being numbered according to their relative location in space: $z$-direction first, then $y$ and finally $x$. An example is shown in Fig. \ref{fig:load_bal}. More complex pre-processing algorithms and load balancing tools will be the focus of future work.

\begin{table}[!ht]
    \centering \small
    \begin{tabular}{c|c|c|c|c|c|c}
        Rank & 
        0 & 1 & 2 & 3 & 4 & 5 \\
        \hline
        Number of local Voronoi cells & 
        17 & 17 & 17 & 17 & 16 & 16 \\
        \hline
        Local Voronoi cell indicies  &
        1-17 & 18-34 & 35-51 & 52-68 & 69-84 & 85-100\\
    \end{tabular}
    \caption{Example patitioning of a mesh with 100 Voronoi cells across 6 processes.}
    \label{tab:partition}
\end{table}

%\begin{table}[!ht]
%    \centering
%    \begin{tabular}{c|c|c}
%        Rank & Number of local Voronoi cells & Local Voronoi cell indicies  \\
%        \hline
%        0 & 17 & 1-17 \\
%        1 & 17 & 18-34\\
%        2 & 17 & 35-51\\
%        3 & 17 & 52-68\\
%        4 & 16 & 69-84\\
%        5 & 16 & 85-100\\
%    \end{tabular}
%    \caption{Example patitioning of a mesh with 100 Voronoi cells across 6 processes.}
%    \label{tab:partition}
%\end{table}

\begin{figure}[!ht]
    \centering
    \includegraphics[width=0.495\textwidth]{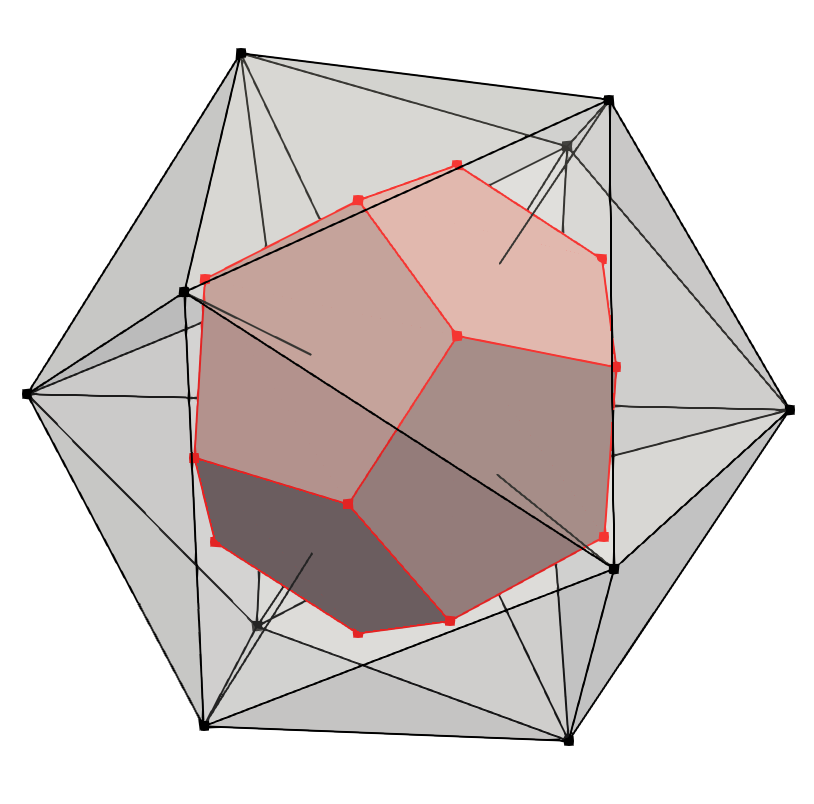}
    \caption{An example of the additional components from the tetrahedral mesh (grayscale) for a single Voronoi cell sub-complex (red).}
    \label{fig:local_cells}
\end{figure}

The result of the partitioning is that each process contains ideally one sub-complex, possibly more, of the whole. Each local sub-complex is augmented with ghost cells on the boundary, specifically where there is an interface to cells on an adjacent process, enabling both primal and dual calculations to be performed  entirely locally. These are the only components which are updated with boundary conditions or during parallel communication exchanges. An example of the additional components from the tetrahedral mesh for a single Voronoi cell sub-complex is shown in Fig. \ref{fig:local_cells}.

%---------------------------------------------------------------------------------------------------------------------------------------
\subsection{Numerical Implementation in ParaGEMS}
\label{subSEC_Numerical_Implementation}
%---------------------------------------------------------------------------------------------------------------------------------------

Applying discrete exterior calculus to the governing equations \eqref{Fourier_EXT} and \eqref{Conservation_EXT}, the resulting system of linear equations modeling thermal diffusion are
\begin{equation}\label{eq:thermal_system}
\left[\begin{array}{cc}
	\frac{1}{\kappa}\star & d^T \\
	d & 0
\end{array}\right]
\left[\begin{array}{cc}
	\omega \\
	\alpha
\end{array}\right] = 
\left[\begin{array}{cc}
	0 \\
	0
\end{array}\right],
\end{equation}
where $\kappa$ has been extracted from the Hodge star and written separately. Dirichlet boundary conditions are applied by subtracting the following vector from the left-hand side of the system of equations
\begin{equation}\label{eq:dirichletRHS}
    \left[\begin{array}{cc}
	d_{\text{BC}}^T T_0 \\
	0
\end{array}\right] \quad \text{ on }\partial M_D,
\end{equation}
where $d_{\text{BC}}^T$ is a diagonal matrix with the negative row sums of $d^T$, marking the end location of incomplete dual edges at the surface of the geometry. In contrast, Neumann boundary conditions are applied by subtracting the following vector from the left-hand side of the system of equations
\begin{equation}\label{eq:neumannRHS}
    \left[\begin{array}{cc}
	-f_0 \\
	d f_0
\end{array}\right] \quad \text{ on }\partial M_N,
\end{equation}
followed by zeroing the rows and columns associated with $\partial M_N$ and placing a unit value on the corresponding diagonal entry, a change of at most 5 values (3 if making use of symmetry). This strategy is equivalent to removing the column and rows associated with the tetrahedral faces in $\partial M_N$, but without altering the form of the system matrix. The process is also used to introduce new cracked (insulated) faces during a simulation. 

Observe that the system matrix in equation \eqref{eq:thermal_system} is in the form
\begin{equation}\label{eq:schur_form}
\left[\begin{array}{cc}
	A & B \\
	B^T & 0
\end{array}\right],
\end{equation}
which is suitable for application of the generalized Schur complement. Therefore, the system matrix and right-hand side are stored as partitioned sparse PETSC matrices and vectors distributed across the available process. A composite Schur preconditioner is then applied along with the flexible generalized minimal residual (FGMRES) iterative method to solve the linear system. The computation of the Schur complement requires the inverse of the $A$ matrix. A common approximation is to use only the diagonal component of $A$; however, in our case the $A$ matrix is already diagonal which should maximize both accuracy and efficiency. The preconditioner for the Schur complement is computed using the Hypre Parasails solver \cite{hypre,parasails}. Additional information regarding the solution process can be found in the software documentation.

%---------------------------------------------------------------------------------------------------------------------------------------
\subsection{I/O and Visualization}
\label{subSEC_IO_and_Visualisation}
%---------------------------------------------------------------------------------------------------------------------------------------

I/O in ParaGEMS is either handled by the root process, which scatters and gathers data to the appropriate adjacent processes, or through built-in PETSc routines. The input Delaunay tetrahedral mesh must be in Tetgen format and contain boundary information (generated using the Tetgen flags \texttt{-f} and \texttt{-e}).  Simulation data is output for post-processing in MATLAB format using built-in PETSc routines, as well as in ASCII VTK format through custom routines for visualization with ParaView. Vector data output in the VTK files is computed at cell barycenters using Whitney interpolation (See software documentation for more information).

%---------------------------------------------------------------------------------------------------------------------------------------
\subsection{Simulation details}
\label{subSEC_Cracks_Introduction}
%---------------------------------------------------------------------------------------------------------------------------------------   

We consider a standard problem of thermal conduction in a cubic domain of unit size, possessing intrinsic thermal conductivity $\kappa$. Dirichlet boundary conditions are prescribed on two opposite sides: temperatures 0 and 1. Neumann boundary conditions are prescribed on the remaining four sides: zero flux density (insulated faces). With this setup, the macroscopic thermal conductivity of the domain, referred here as the effective (apparent) thermal conductivity and denoted as $\kappa_e$, equals numerically the calculated flux density through either of the sides with Dirichlet boundary conditions. In the case when the thermal conductivity prescribed to dual faces is constant across the domain (mimicking continuous and homogeneous material situation), the effective thermal conductivity equals the prescribed value. It will be demonstrated that this remains true irrespective of the cell complex used to cover the domain. In the case of variable thermal conductivity across dual faces, the effective thermal conductivity will be a function of the particular distribution of local conductivities, mimicking material heterogeneity.

To demonstrate use of the software to study the effect of emerging discontinuities in the material on its thermal conductivity, we consider the following. A discontinuity or a crack in the material is an internal surface that prohibits heat transfer across. The heat flux in our formulation is across faces of the cell complex selected as dual (a 2-form on the dual). An emergence of a crack is therefore equated to a change of the thermal conductivity of a selected face from its current value to zero.  

For comparison with available experimental data, we have devised two scenarios for crack introduction. In both cases, we use a sequence of steady-state solutions of the system, starting with a domain with no cracks, introducing one single crack at each step and calculating the resulting effective thermal conductivity. The process terminates when the domain becomes non-conductive due to specific arrangement of the crack population. 

In the first scenario, referred to as deterministic, the selection of a face to be converted to crack is based on the solution from the previous step: the face with highest flux density is converted to a crack. In the second scenario, referred to as stochastic, cracks are introduced by a random selection of faces for which the currently prescribed thermal conductivity is changed to zero. The second scenario provides the effective thermal conductivity as a function of either the number or the volume density of randomly distributed cracks. This would allow for comparison with experimental data of thermal conductivity for any given number of volume density of cracks, assuming the cracks are randomly distributed. The first scenario is intended to represent a maximal thermal `damage', i.e. a situation where the emerging cracks are reducing the effective conductivity maximally. This would allow for comparison with experimental data of thermal conductivity where the cracks are introduced by some thermo-mechanical process that leads to non-uniform or ordered distribution.

%---------------------------------------------------------------------------------------------------------------------------------------   
\section{Results and Discussion}
\label{Sec_Results_and_Discussion}
%--------------------------------------------------------------------------------------------------------------------------------------- 

 ParaGEMS was run using a local high-performance computing cluster, the specifications of which are detailed below.  For users' convenience, scripts used to run ParaGEMS on either a laptop computer or a high-performance computing cluster are provided within the repository.   A full list of ParaGEMS input parameters, along with an explanation of each, is also documented.

\subsection{Baseline verification}

\begin{figure}
    \centering
    \includegraphics[width=0.75\textwidth]{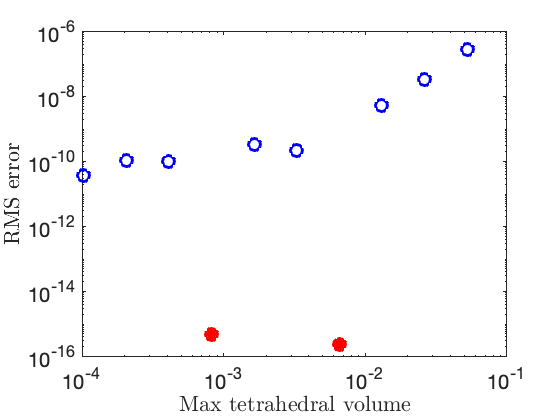}
    \caption{Volume integrated RMS error as a function of maximum Delaunay tetrahedron volume (For reference, the full geometry has unit volume). }
    \label{fig:accuracy}
\end{figure}

Baseline verification is obtained from thermal diffusion in a pristine cube, which has a linear analytic solution as a function distance through the cube along the direction of heat flow. The accuracy of ParaGEMS as a function of maximum tetrahedron volume is shown in Fig. \ref{fig:accuracy}. The accuracy is is evaluated as the RMS error
\begin{equation}
    \text{RMS error} = \sqrt{\sum_{\forall\  \text{tetrahedra}} \left(T_{i,\text{comp}} - T_{i,\text{exact}}\right)^2 V_i}.
\end{equation}

Most meshes used in this study contained dual edges with zero or near-zero edge lengths. To improve conditioning, Hodge-star entries containing these lengths are limited to a value of $10^{-8}$. The influence of this limiter is visible in the majority of results as non-machine-zero error. The downward trend observed as the mesh is refined is due to the percentage of limited values decreasing. To confirm this is the likely cause of the error, the absolute and relative tolerances were reduced from $10^{-10}$ to $10^{-14}$, however no change in the result was observed. Furthermore, two particular meshes, with results highlighted in red, had sufficiently long dual edge lengths to avoid the interference of the limiter. In these cases, the error is of the order of machine precision. This also indicates that for a mesh of sufficient quality, the method is exact for linear solutions.

%++++++++++++++++++++++++++++++++++++++++++++++++++++++++++++++++++++++++++++++++++++++++++++

%++++++++++++++++++++++++++++++++++++++++++++++++++++++++++++++++++++++++++++++++++++++++++++

\subsection{Parallel scaling}

Both strong and weak scaling are evaluated on a local high-performance computing cluster at the University of Manchester, UK,  with hardware consisting of nodes of 2 x 16-core Intel Skylake Gold 6130 CPUs at 2.10 GHz and 192 GB RAM connected through a Mellanox MT27800 100 Gb/s InfiniBand interconnect. This study is done with a series of successively finer meshes  created by halving the maximum tetrahedron volume on each level, resulting in meshes with approximately 305 thousand tetrahedra to 18.7 million. In each case, the number of unique triangular faces is about double the number of tetrahedra. Therefore, the number of equations solved is between 928 thousand and 56 million. 

\begin{figure}
    \centering
    \includegraphics[width=0.75\textwidth]{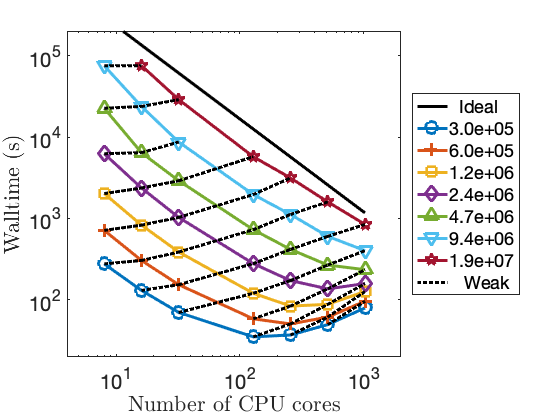}
    \caption{Strong and week scaling performance of ParaGEMS combined with the solution obtained with PETSc on 8 to 1024 Skylake cores.}
    \label{fig:scalePETSc}
\end{figure}

Results of the study are shown in Fig. \ref{fig:scalePETSc}, and indicate excellent strong scaling of the ParaGEMS library up to the maximum number of cores on the three largest meshes, where computation effort overshadows the parallel communication overhead. On coarser meshes, the cost of parallelization becomes an increasingly significant portion of the overall cost, diminishing the advantage of adding more processors. On the three coarsest grids, the cost of parallelization eventually begins to increase the overall solution time. While strong scaling is very good, the weak scaling has room for improvement.

\subsection{Deterministic cracking process}
\label{subSec_Deterministic_cracking_process}
\begin{figure}
  \centering
    \includegraphics[width=0.49\textwidth]{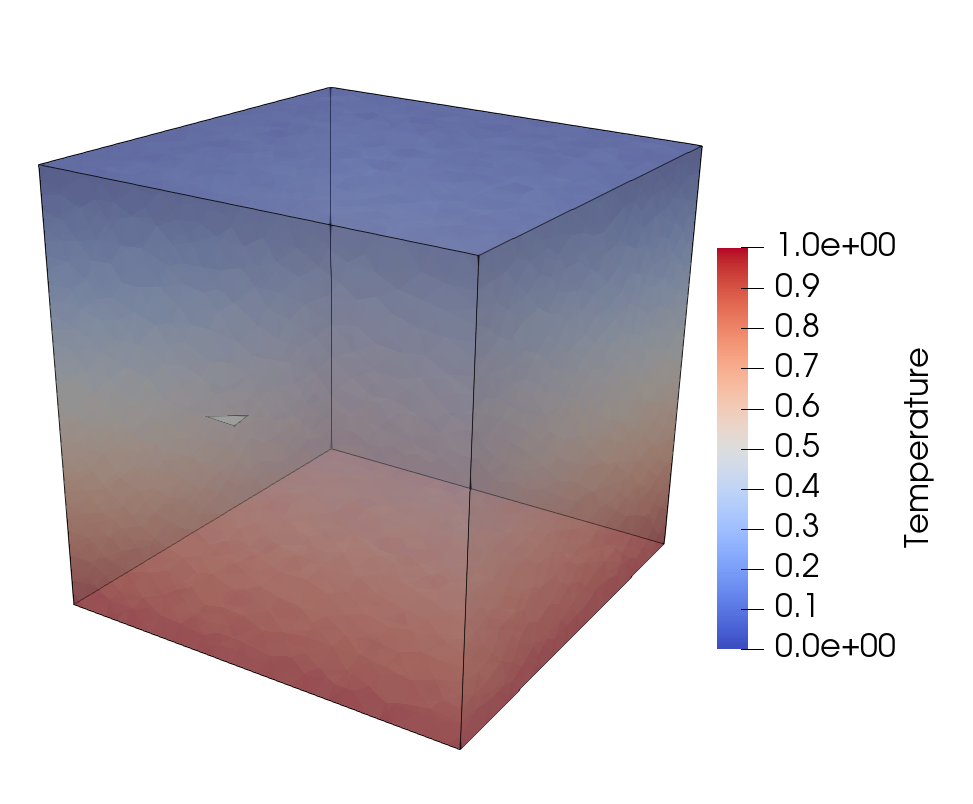}
    \includegraphics[width=0.49\textwidth]{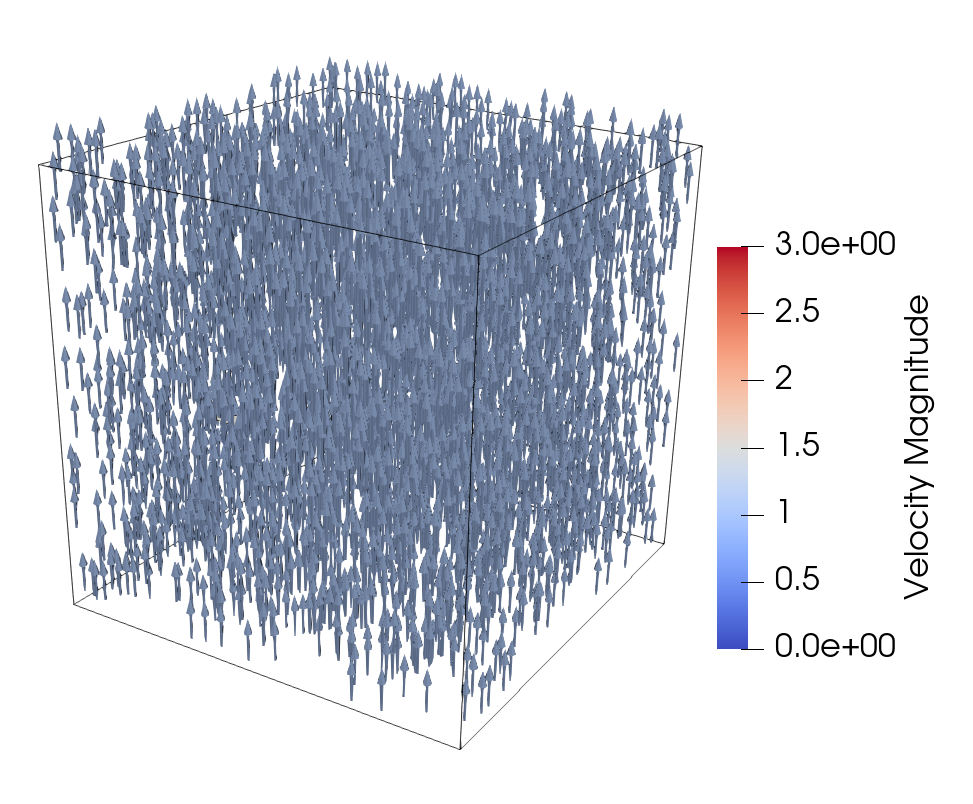}\\
    \includegraphics[width=0.49\textwidth]{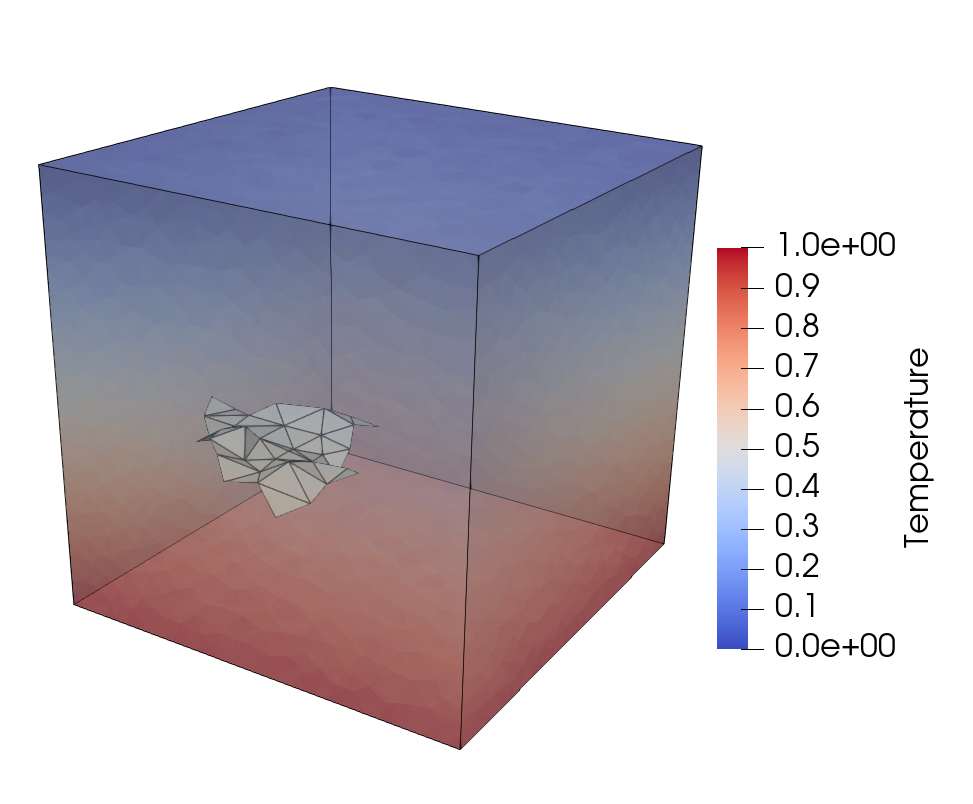}
    \includegraphics[width=0.49\textwidth]{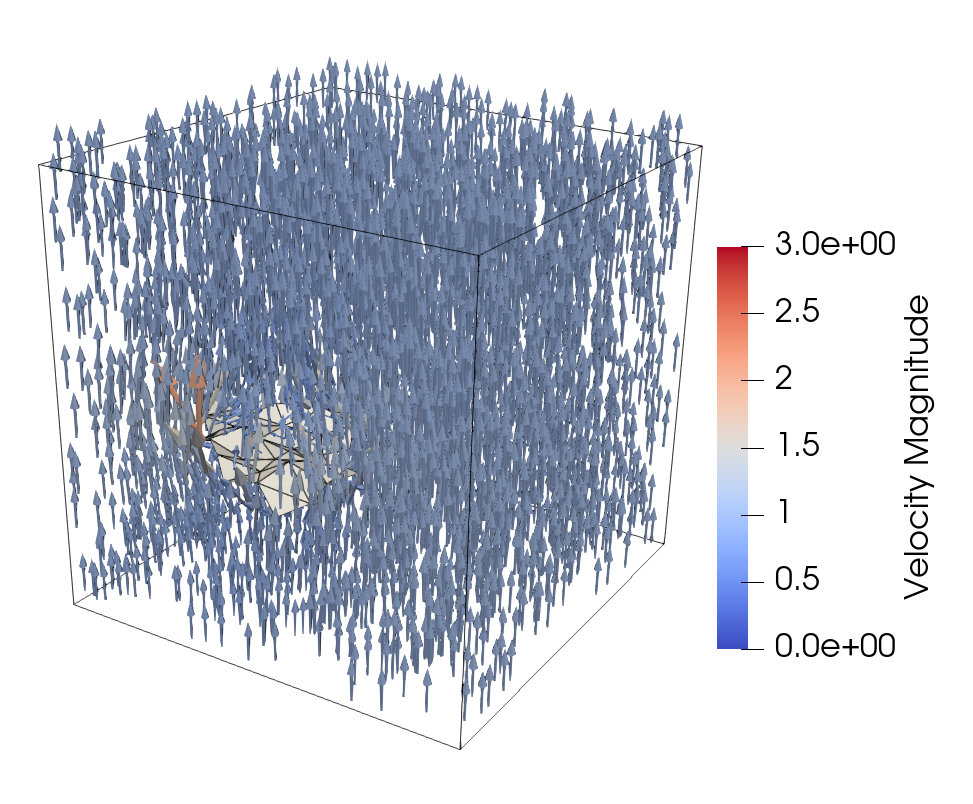}\\
    \includegraphics[width=0.49\textwidth]{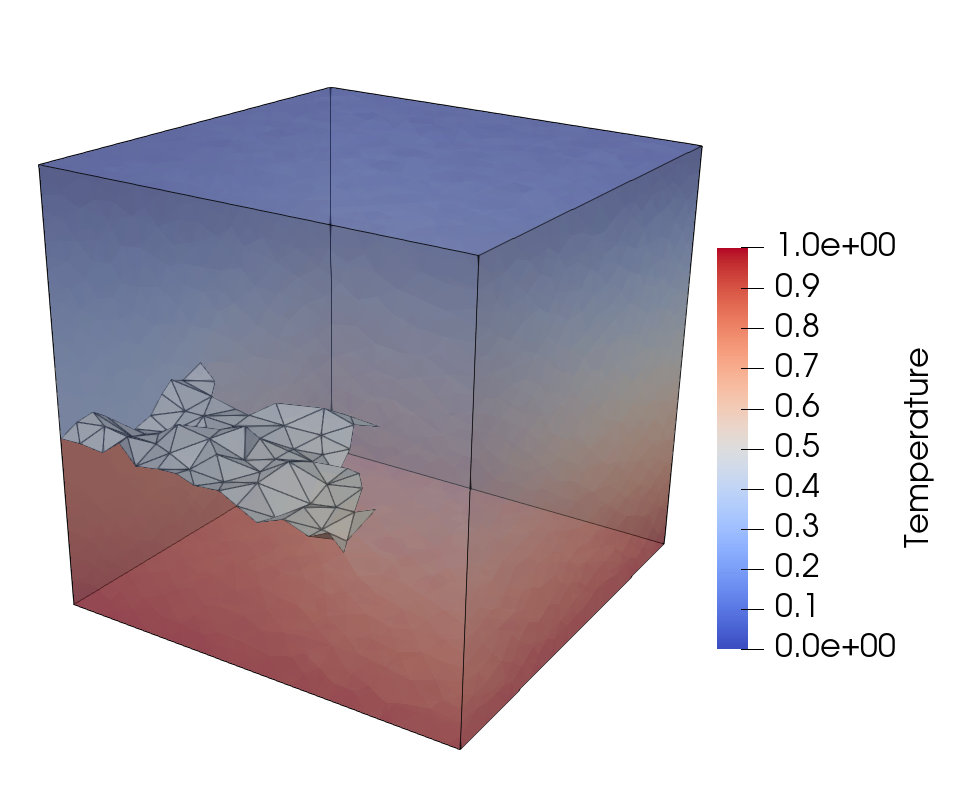}
    \includegraphics[width=0.49\textwidth]{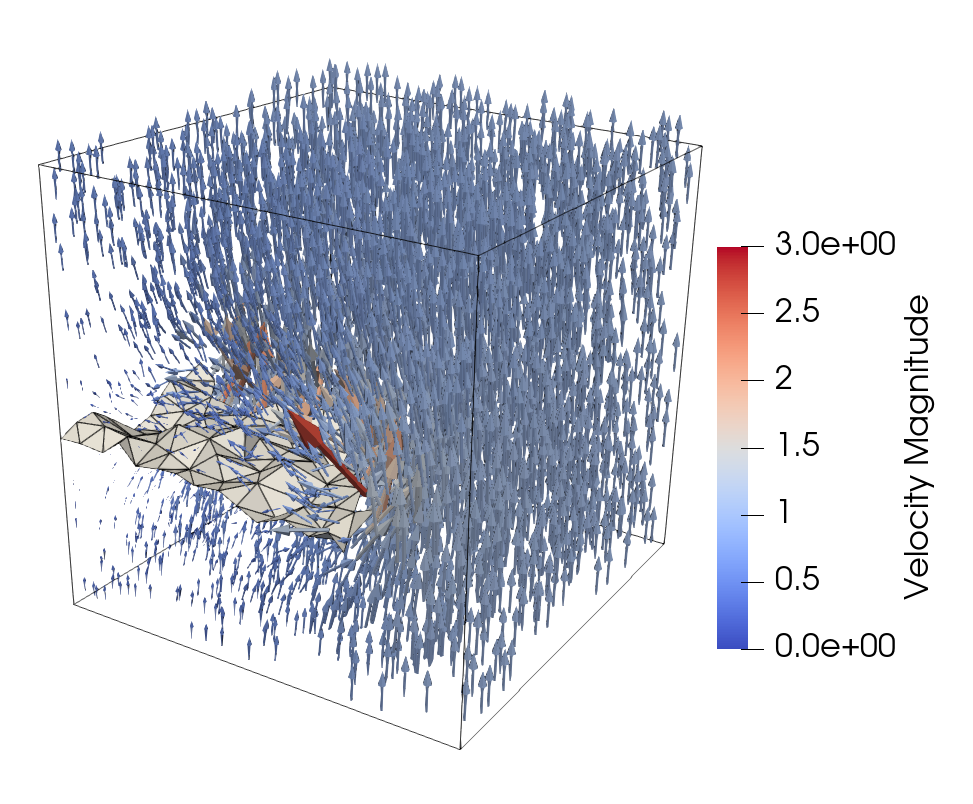}\\[-6ex]
    \includegraphics[width=0.15\textwidth]{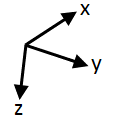} 
        \caption{Visualization of deterministic cracking process: single crack (row 1); 10\% crack fraction (row 2); and 25\% crack fraction (row 3).} 
    \label{paraview_det}
\end{figure}

\begin{figure}
  \centering
    \includegraphics[width=0.49\textwidth]{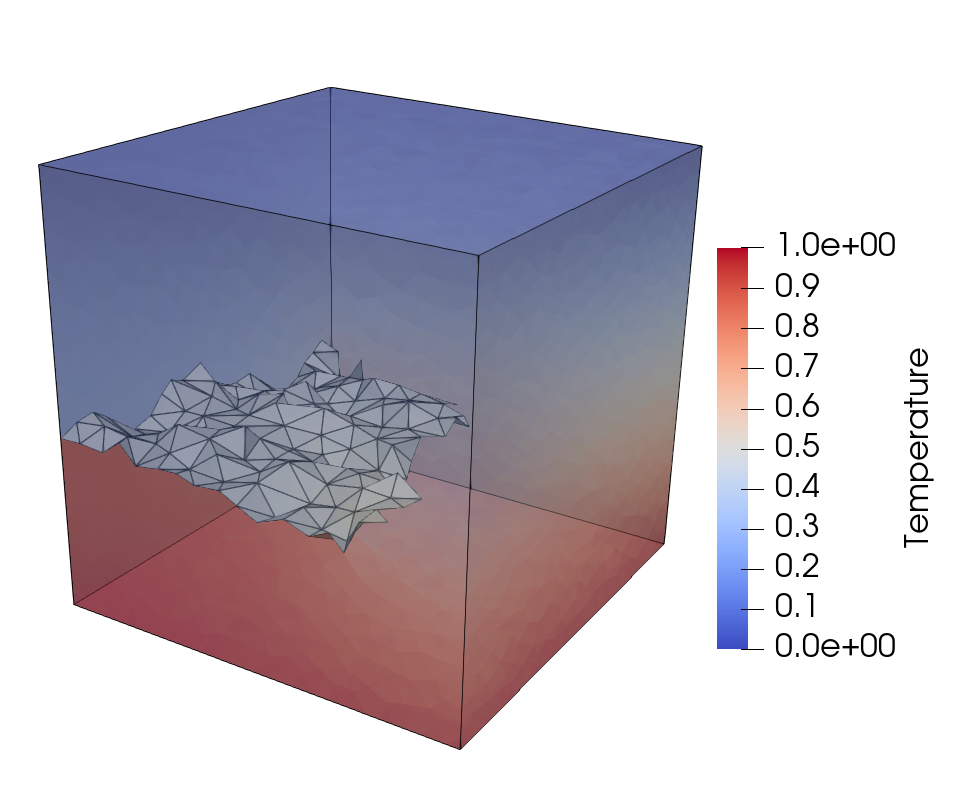}
    \includegraphics[width=0.49\textwidth]{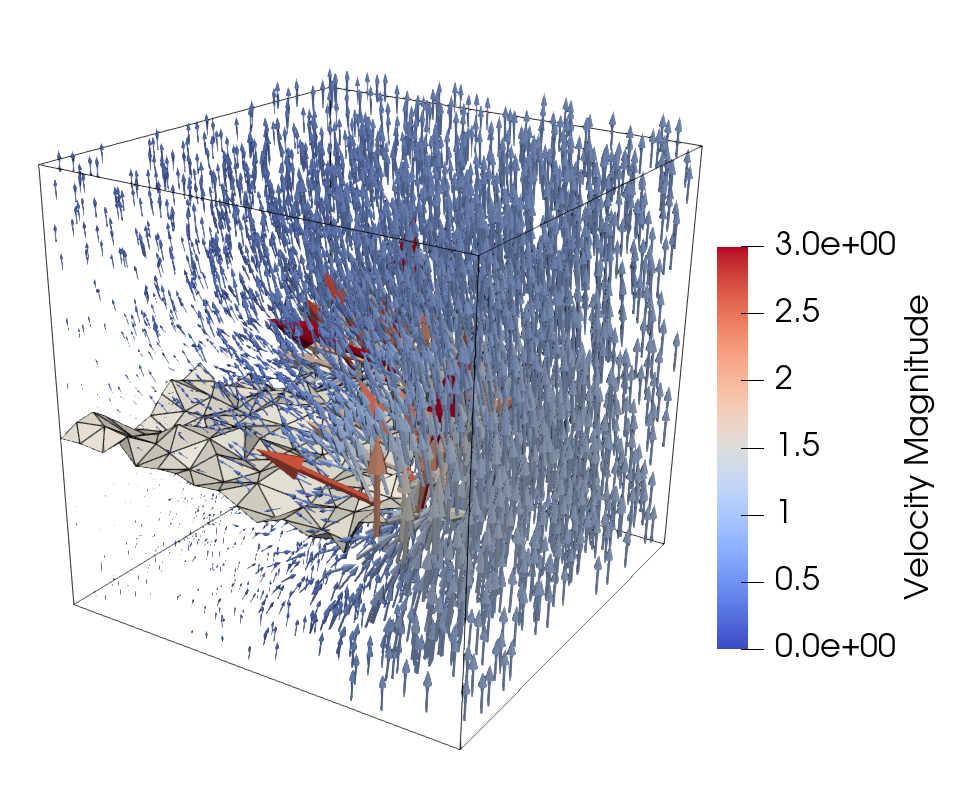}\\
    \includegraphics[width=0.49\textwidth]{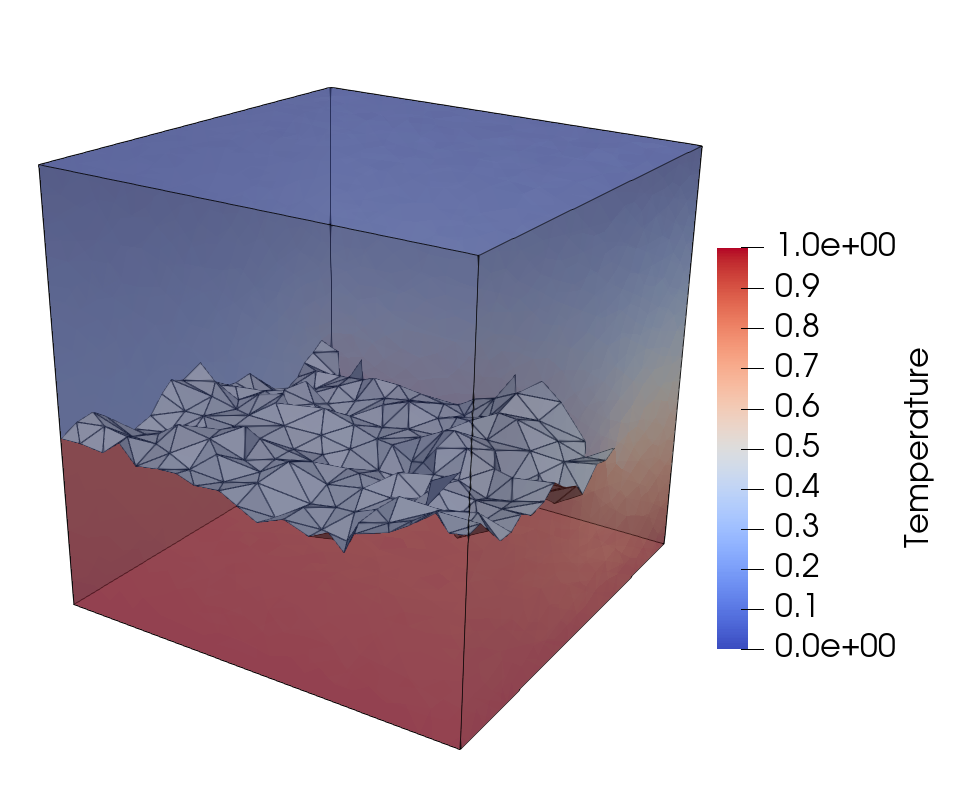}
    \includegraphics[width=0.49\textwidth]{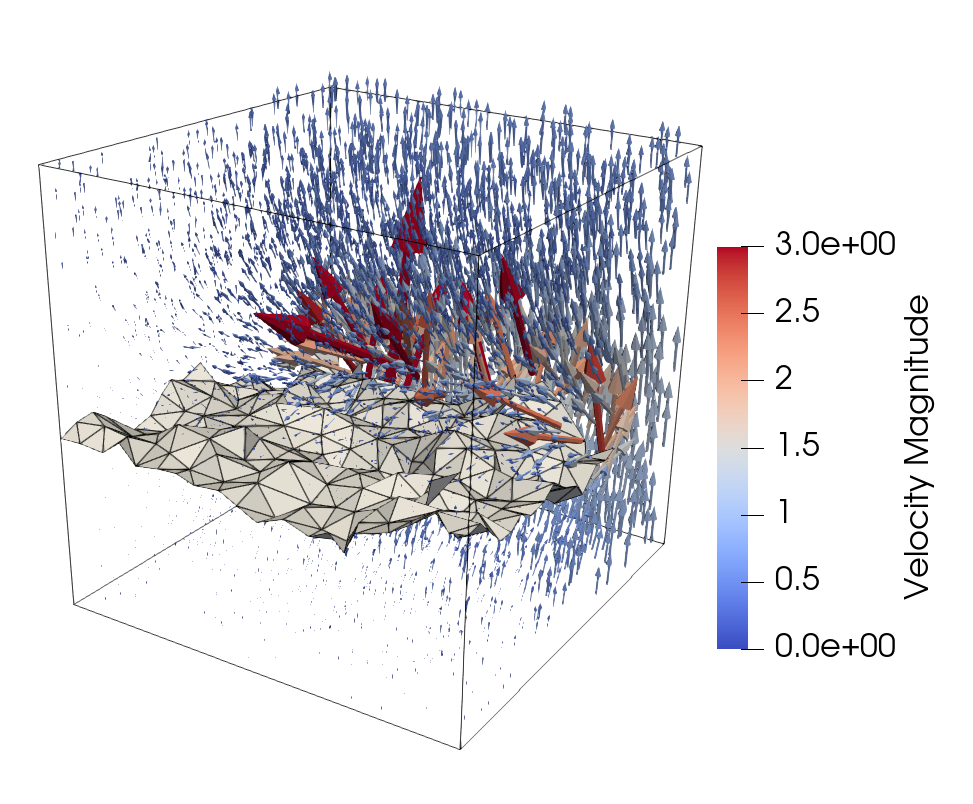}\\
    \includegraphics[width=0.49\textwidth]{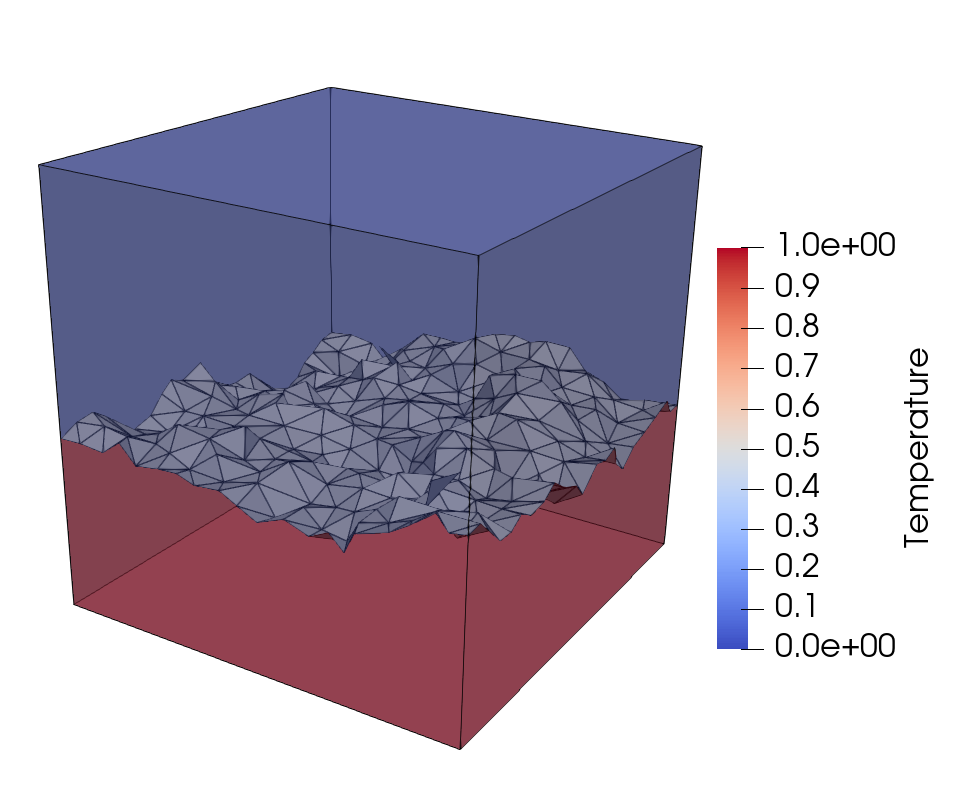}
    \includegraphics[width=0.49\textwidth]{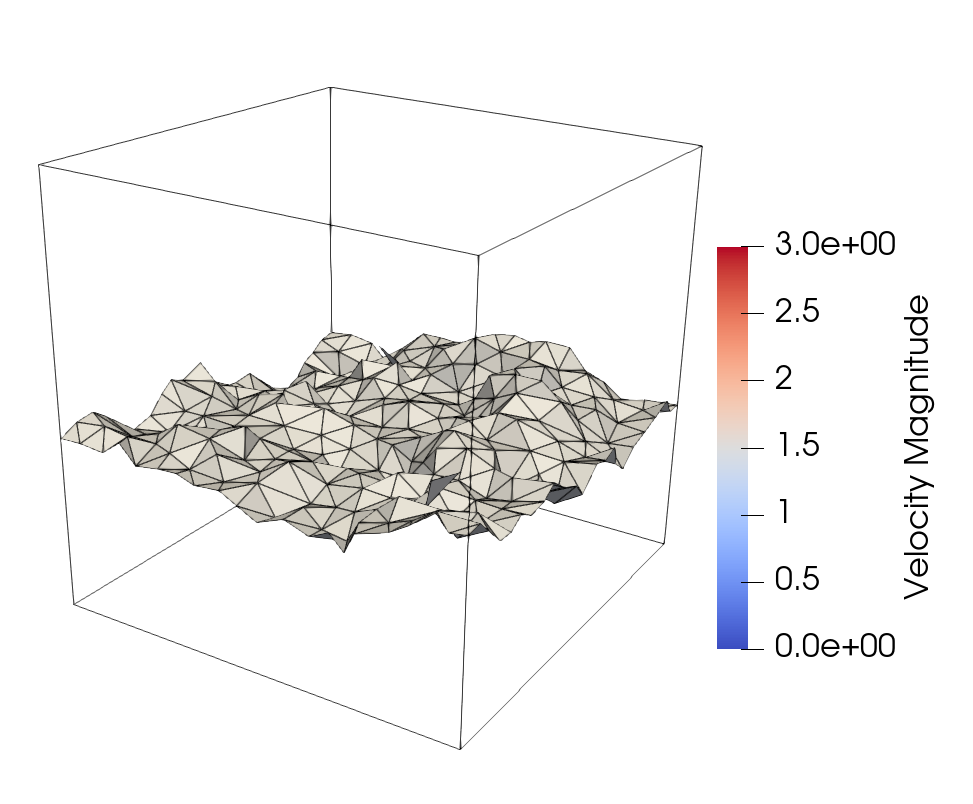}\\[-6ex]
    \includegraphics[width=0.15\textwidth]{AXIS1.png} 
        \caption{Visualization of deterministic cracking process: 50\% crack fraction (row 1); 75\% crack fraction (row 2); and 100\% crack fraction (row 3).} 
    \label{paraview_det1}
\end{figure}

%--------------------------------------------------------------------------------------------------------------------------------------- 
A unit cube is tessellated into a tetrahedral mesh containing 44879 triangular faces of which 39667 are non-boundary faces. At the start of the simulations, all non-boundary faces are assigned equal conductivity $\kappa \ne 0$. After each simulation step, the face with maximal heat flux is cracked, i.e. its thermal conductivity is set to zero, and $\kappa_e$ of the updated domain is calculated. The process is repeated until $\kappa_e=0$, i.e. the domain becomes thermally non-conductive, producing $\kappa_e$ as a function of cracked faces. Figs.~\ref{paraview_det} and~\ref{paraview_det1} show the solution at several crack configurations, selected as fractions of the final crack which makes the domain non-conductive. The left- and right-hand columns show the temperature distribution and the heat velocity vectors through faces, respectively. The uppermost row of Fig.~\ref{paraview_det} shows the first (single) crack that emerged nearly perpendicularly to the heat flow direction, and the velocity vectors parallel to the flow ($z$) direction. The effect of crack growth on the temperature distribution and heat velocity can be clearly followed and quantified.

To illustrate the dependence of $\kappa_e$ on cracking, we introduce a count-based damage parameter, $D_n$, defined as the ratio between number of cracked faces and the total number of non-boundary faces. The calculated relationship is shown by the top graph in Fig.~\ref{100MC}.

%--------------------------------------------------------------------------------------------------------------------------------------- 
\subsection{Stochastic cracking process}
\label{subSec_Stochastic_cracking_process}
%---------------------------------------------------------------------------------------------------------------------------------------  

\begin{figure}
\begin{center}
\includegraphics[width=0.75\textwidth]{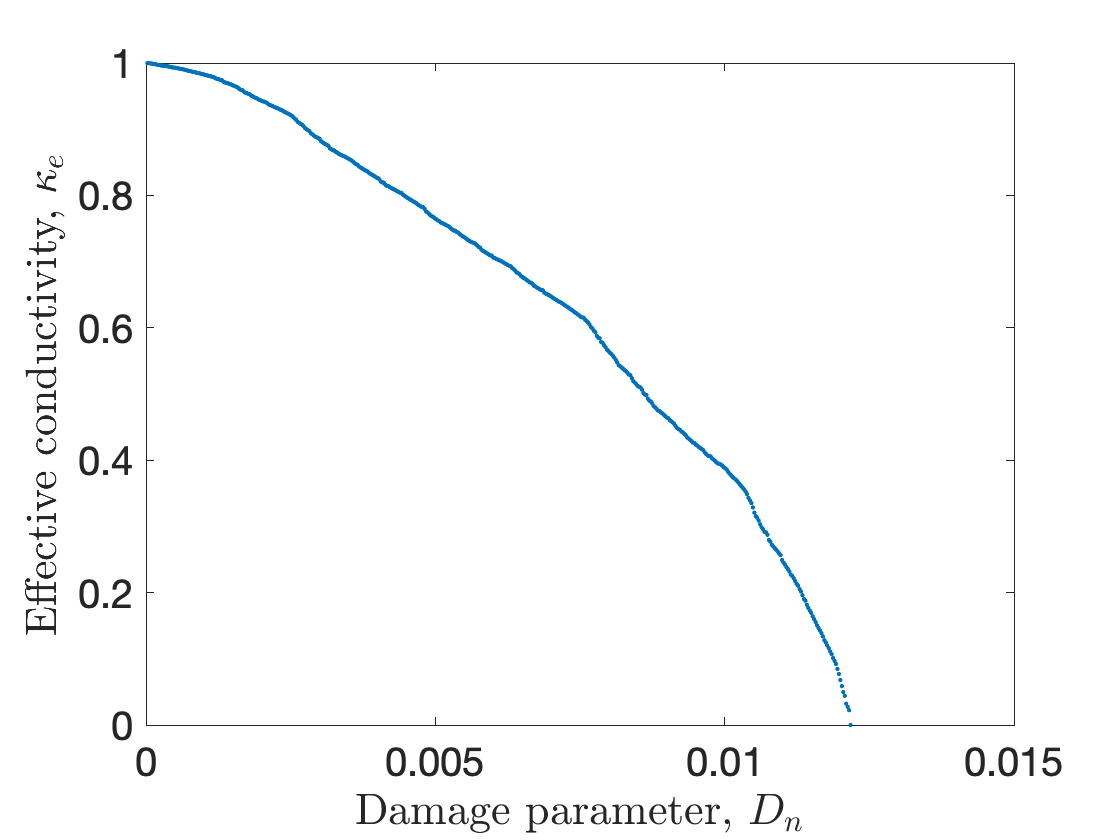}
\includegraphics[width=0.75\textwidth]{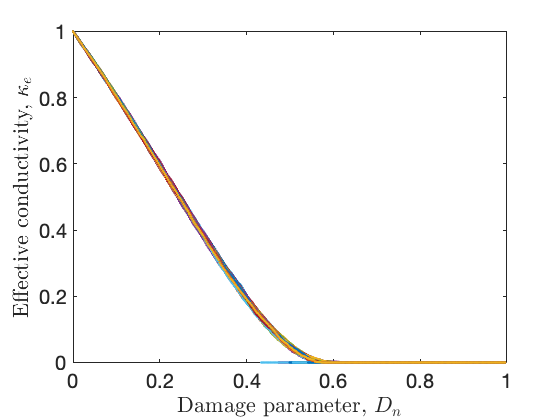}
 \caption{Calculated dependence of $\kappa_e$ on the damage parameter: (top) by the deterministic cracking process; and (bottom) by the stochastic cracking process for 100 Monte Carlo paths.}
\label{100MC}
 \end{center}
\end{figure}

The process of stochastic cracking is simulated on the tetrahedral mesh with the initial assignment of conductivity to non-boundary faces used for the deterministic cracking process. The analysis is based on the well-known Monte Carlo method. After each simulation step, a randomly selected non-boundary face is cracked, and $\kappa_e$ of the updated domain is calculated. The process is repeated until $\kappa_e=0$, producing $\kappa_e$ as a function of cracked faces. This constitutes a single Monte Carlo path. The process can be repeated $N$ times, representing $N$ Monte Carlo trials.  Fig.~\ref{100MC} (bottom) shows $\kappa_e$ plotted against the damage parameter, for all Monte Carlo paths.  The mean effective thermal conductivity has not been plotted, since the variance of $\kappa_e$ observed in the 100 Monte Carlo trials, is small.  The error in a Monte Carlo model scales inversely as a function of $\sqrt{N}$. The effective thermal conductivity for certain Monte Carlo paths appear to become abruptly zero, observed for crack fractions between $\approx $0.4 and $\approx$0.6. This arises from stochastic cracking paths by chance following a crack sequence with deterministic-like behaviour. In reality, the set of all distinct Monte Carlo paths includes both the deterministic result and a path which prioritizes the cracking of faces most parallel to the flow, which have little or no impact on the effective thermal conductivity, as well as everything in between.

The initial conditions (geometry and conductivity distribution) are the same for both graphs in Fig.~\ref{100MC}, allowing for direct comparison of $\kappa_e$ as a function of the damage parameter. Remarkable differences exist: the deterministic cracking process makes the domain non-conductive at damage parameter of $\approx 0.012$, which is considerably less than that for the stochastic process of  $\approx$0.55. One can consider the deterministic process as close to the worst-case scenario for the reduction of thermal conductivity with cracking. The stochastic cracking process does not represent the best-case scenario (which could be realized by firstly cracking all faces nearly parallel to the heat flux), but could be a physically realistic representation of damage introduced by bulk thermal shock. 

%++++++++++++++++++++++++++++++++++++++++++++++++++++++++++++++++++++++++++++++++++++++++++++
%++++++++++++++++++++++++++++++++++++++++++++++++++++++++++++++++++++++++++++++++++++++++++++

%++++++++++++++++++++++++++++++++++++++++++++++++++++++++++++++++++++++++++++++++++++++++++++
%++++++++++++++++++++++++++++++++++++++++++++++++++++++++++++++++++++++++++++++++++++++++++++ 

%--------------------------------------------------------------------------------------------------------------------------------------------------------------
\section{Conclusions and perspectives}
\label{Sec_Conclusions_and_perspectives}
%--------------------------------------------------------------------------------------------------------------------------------------------------------------

We have presented development and capabilities of a parallelized software based on discrete exterior calculus for solving 3D elliptic boundary value problems. The portability of the code means that execution is feasible on distributed memory HPC architectures. One particular advantage of the present approach is that since the discrete algebraic equations formulated via DEC are invariant through the embedding of each simplex and of the entire mesh, the present method can be applied to simplicial approximations of non-flat manifolds \cite{Mamdouh_2018}. Furthermore, as demonstrated in this article, it provides an effortless means to introduce of strong heterogeneities and discontinuities.

Future work includes extending the code to fluid and solid mechanics, and use of a parallel DEC library for musical isomorphisms, interpolation, Hodge decomposition and wedge products. It is anticipated that performance will increase further through the optimal use of ARM and GPU architectures. Finally, a virtual reality graphical user interface for visualization of meshes and discontinuous permeability represents an exciting future goal.

%--------------------------------------------------------------------------------------------------------------------------------------------------------------
\section*{Acknowledgements}
\label{Acknowledgements}
%--------------------------------------------------------------------------------------------------------------------------------------------------------------
The authors are grateful to Professor Anil Hirani from the University of Illinois Urbana-Champaign for the many fruitful discussions regarding the theory and application of discrete exterior calculus. The authors acknowledge the financial support for this work by the Engineering and Physical Sciences Research Council, UK (EPSRC) via grant EP/N026136/1.  Furthermore, the authors would like to acknowledge the assistance given by Research IT and the use of the Computational Shared Facility at The University of Manchester.

%------------------------------------------------------------------------------------------------------------------  
%  References   
%------------------------------------------------------------------------------------------------------------------  
%\section*{References}
%\bibliographystyle{model1-num-names}
%\bibliographystyle{plainnat}
\bibliography{Odybib}

\begin{thebibliography}{10}
\expandafter\ifx\csname url\endcsname\relax
  \def\url#1{\texttt{#1}}\fi
\expandafter\ifx\csname urlprefix\endcsname\relax\def\urlprefix{URL }\fi
\expandafter\ifx\csname href\endcsname\relax
  \def\href#1#2{#2} \def\path#1{#1}\fi

\bibitem{Lewis_FEM_1996}
R.~Lewis, K.~Morgan, H.~Thomas, K.~Seetharamu, Finite Element Method in Heat
  Transfer Analysis, Wiley-Blackwell, 1996.

\bibitem{Ibanez_BEM_2002}
M.-T. Ibanez, H.~Power, Advanced Boundary Elements for Heat Transfer, WIT
  Press, 2002.

\bibitem{Lu_JMPS_1995}
T.~Lu, J.~Hutchinson, Thermal conductivity and expansion of cross-ply
  composites with matrix cracks, Journal of the Mechanics and Physics of Solids
  43~(8) (1995) 1175--1198.
\newblock \href
  {http://dx.doi.org/https://doi.org/10.1016/0022-5096(95)00033-F}
  {\path{doi:https://doi.org/10.1016/0022-5096(95)00033-F}}.

\bibitem{Pestchanyi_FED_2010}
S.~Pestchanyi, I.~Garkusha, I.~Landman, Simulation of tungsten armour cracking
  due to small elms in iter, Fusion Engineering and Design 85~(7) (2010)
  1697--1701, proceedings of the Ninth International Symposium on Fusion
  Nuclear Technology.
\newblock \href
  {http://dx.doi.org/https://doi.org/10.1016/j.fusengdes.2010.05.005}
  {\path{doi:https://doi.org/10.1016/j.fusengdes.2010.05.005}}.

\bibitem{Shen_CBM_2015}
L.~Shen, Q.~Ren, N.~Xia, L.~Sun, X.~Xia, Mesoscopic numerical simulation of
  effective thermal conductivity of tensile cracked concrete, Construction and
  Building Materials 95 (2015) 467--475.

\bibitem{Shen_CBM_2017}
L.~Shen, Q.~Ren, L.~Zhang, Y.~Han, G.~Cusatis, Experimental and numerical study
  of effective thermal conductivity of cracked concrete, Construction and
  Building Materials 153 (2017) 55--68.
\newblock \href {http://dx.doi.org/10.1016/j.conbuildmat.2017.07.038}
  {\path{doi:10.1016/j.conbuildmat.2017.07.038}}.

\bibitem{Simionato_ZAMM_2019}
F.~Simionato, C.~Daros, Boundary element method analysis for mode iii linear
  fracture mechanics in anisotropic and nonhomogeneous media, ZAMM - Journal of
  Applied Mathematics and Mechanics / Zeitschrift für Angewandte Mathematik
  und Mechanik 99~(10) (2019) e201800211.
\newblock \href {http://dx.doi.org/10.1002/zamm.201800211}
  {\path{doi:10.1002/zamm.201800211}}.

\bibitem{Oschmann_JPE_2016}
T.~Oschmann, M.~Schiemann, H.~Kruggel-Emden, Development and verification of a
  resolved 3d inner particle heat transfer model for the discrete element
  method (dem), Powder Technology 291 (2016) 392--407.
\newblock \href {http://dx.doi.org/10.1016/j.powtec.2015.12.008}
  {\path{doi:10.1016/j.powtec.2015.12.008}}.

\bibitem{Peng_PECS_2020}
Z.-B. Peng, E.~Doroodchi, B.~Moghtaderi, Heat transfer modelling in discrete
  element method (dem)-based simulations of thermal processes: Theory and model
  development, Progress in Energy and Combustion Science 79 (2020) 100847.
\newblock \href {http://dx.doi.org/10.1016/j.pecs.2020.100847}
  {\path{doi:10.1016/j.pecs.2020.100847}}.

\bibitem{Clearly_JCP_1999}
P.~Cleary, J.~Monaghan, Conduction modelling using smoothed particle
  hydrodynamics, Progress in Energy and Combustion Science 148~(1) (1999)
  227--264.
\newblock \href {http://dx.doi.org/10.1006/jcph.1998.6118}
  {\path{doi:10.1006/jcph.1998.6118}}.

\bibitem{Alshaer_CMS_2017}
A.~Alshaer, B.~Rogers, L.~Li, Smoothed particle hydrodynamics (sph) modelling
  of transient heat transfer in pulsed laser ablation of al and associated
  free-surface problems, Computational Materials Science 127 (2017) 161--179.
\newblock \href {http://dx.doi.org/10.1016/j.commatsci.2016.09.004}
  {\path{doi:10.1016/j.commatsci.2016.09.004}}.

\bibitem{Bobaru_JCP_2012}
F.~Bobaru, M.~Duangpanya, A peridynamic formulation for transient heat
  conduction in bodies with evolving discontinuities, Journal of Computational
  Physics 213~(7) (2012) 2764--2785.
\newblock \href {http://dx.doi.org/10.1016/j.jcp.2011.12.017}
  {\path{doi:10.1016/j.jcp.2011.12.017}}.

\bibitem{Yan_JH_2021}
H.-X. Yan, M.~Sedighi, A.~Jivkov, Peridynamics modelling of coupled water flow
  and chemical transport in unsaturated porous media, Journal of Hydrology in
  press (2021) xxxx--xxxx.
\newblock \href {http://dx.doi.org/10.1016/j.jhydrol.2020.125648}
  {\path{doi:10.1016/j.jhydrol.2020.125648}}.

\bibitem{Hirani_DEC_2003}
A.~Hirani, Discrete exterior calculus, Ph.D. thesis, California Institute of
  Technology (2003).
\newblock \href {http://dx.doi.org/10.7907/ZHY8-V329}
  {\path{doi:10.7907/ZHY8-V329}}.

\bibitem{Belayachi_JBE_2019}
N.~Belayachi, C.~Mallet, M.~E. Marzak, Thermally-induced cracks and their
  effects on natural and industrial geomaterials, Journal of Building
  Engineering 25 (2019) 100806.
\newblock \href {http://dx.doi.org/10.1016/j.jobe.2019.100806}
  {\path{doi:10.1016/j.jobe.2019.100806}}.

\bibitem{Schulz_DCG_2020}
E.~Schulz, G.~Tsogtgerel, Convergence of discrete exterior calculus
  approximations for poisson problems, Discrete \& Computational Geometry 63
  (2020) 364--376.
\newblock \href {http://dx.doi.org/10.1007/s00454-019-00159-x}
  {\path{doi:10.1007/s00454-019-00159-x}}.

\bibitem{Arnold_CBMS_2018}
D.~Arnold, Finite Element Exterior Calculus, Vol.~93 of CBMS-NSF Regional
  Conference Series in Applied Mathematics, Society for Industrial and Applied
  Mathematics (SIAM), Philadelphia, PA, 2018.

\bibitem{Vu_IJHMT_2015}
M.~Vu, S.~Nguyen, M.~Vu, A.~Tang, V.~To,
  \href{http://www.sciencedirect.com/science/article/pii/S0017931015006171}{Heat
  conduction and thermal conductivity of 3d cracked media}, International
  Journal of Heat and Mass Transfer 89 (2015) 1119--1126.
\newblock \href
  {http://dx.doi.org/https://doi.org/10.1016/j.ijheatmasstransfer.2015.05.113}
  {\path{doi:https://doi.org/10.1016/j.ijheatmasstransfer.2015.05.113}}.
\newline\urlprefix\url{http://www.sciencedirect.com/science/article/pii/S0017931015006171}

\bibitem{Hirani_IJCMSM_2015}
A.~Hirani, K.~Nakshatrala, J.~Chaudhry, Numerical method for darcy flow derived
  using discrete exterior calculus, International Journal for Computational
  Methods in Engineering Science and Mechanics 16~(3) (2015) 151--169.
\newblock \href {http://dx.doi.org/10.1080/15502287.2014.977500}
  {\path{doi:10.1080/15502287.2014.977500}}.

\bibitem{Si_ACM_2015}
H.~Si, Tetgen, a delaunay-based quality tetrahedral mesh generator, ACM
  Transactions on Mathematical Software 41~(2) (2015) 11.
\newblock \href {http://dx.doi.org/10.1145/2629697}
  {\path{doi:10.1145/2629697}}.

\bibitem{Kit_JEP_1972}
G.~Kit, O.~Poberezhnyi, Determining the steady-state temperature field in a
  cracked plate with heat transfer from the lateral surfaces, Journal of
  Engineering Physics 23 (1972) 894--898.

\bibitem{Glushko_AA_2016}
A.~Glushko, A.~S. Ryabenko, V.~Petrova, E.~A. Loginova, Heat distribution in a
  plane with a crack with a variable coefficient of thermal conductivity,
  Asymptot. Anal. 98 (2016) 285--307.

\bibitem{Hoenig_JCM_1983}
A.~Hoenig, Thermal conductivities of a cracked solid, Journal of Composite
  Materials 17~(3) (1983) 231--237.
\newblock \href {http://dx.doi.org/10.1177/002199838301700304}
  {\path{doi:10.1177/002199838301700304}}.

\bibitem{Savruk_MS_1987}
M.~Savruk, V.~Zelenyak, The plane problem of thermal conductivity and thermal
  elasticity for a finite piecewise uniform body with cracks, Mater. Sci. 23
  (1987) 502--510.
\newblock \href {http://dx.doi.org/https://doi.org/10.1007/BF01148677}
  {\path{doi:https://doi.org/10.1007/BF01148677}}.

\bibitem{Hasselman_JCM_1987}
D.~Hasselman, L.~F. Johnson, Effective thermal conductivity of composites with
  interfacial thermal barrier resistance, Journal of Composite Materials 21~(6)
  (1987) 508--515.
\newblock \href {http://dx.doi.org/10.1177/002199838702100602}
  {\path{doi:10.1177/002199838702100602}}.

\bibitem{Benveniste_JAP_1989}
Y.~Benveniste, T.~Miloh, An exact solution for the effective thermal
  conductivity of cracked bodies with oriented elliptical cracks, Journal of
  Applied Physics 66~(1) (1989) 176--180.
\newblock \href {http://dx.doi.org/10.1063/1.343900}
  {\path{doi:10.1063/1.343900}}.

\bibitem{Hu_IJHTMT_2013}
K.~Hu, Z.~Chen,
  \href{http://www.sciencedirect.com/science/article/pii/S0017931013002391}{Transient
  heat conduction analysis of a cracked half-plane using dual-phase-lag
  theory}, International Journal of Heat and Mass Transfer 62 (2013) 445--451.
\newblock \href
  {http://dx.doi.org/https://doi.org/10.1016/j.ijheatmasstransfer.2013.03.032}
  {\path{doi:https://doi.org/10.1016/j.ijheatmasstransfer.2013.03.032}}.
\newline\urlprefix\url{http://www.sciencedirect.com/science/article/pii/S0017931013002391}

\bibitem{Tran_JAG_2018}
A.~Tran, M.~Vu, S.~Nguyen, T.~Dong, K.~Le-Nguyen,
  \href{http://www.sciencedirect.com/science/article/pii/S0926985116306929}{Analytical
  and numerical solutions for heat transfer and effective thermal conductivity
  of cracked media}, Journal of Applied Geophysics 149 (2018) 35--41.
\newblock \href
  {http://dx.doi.org/https://doi.org/10.1016/j.jappgeo.2017.12.012}
  {\path{doi:https://doi.org/10.1016/j.jappgeo.2017.12.012}}.
\newline\urlprefix\url{http://www.sciencedirect.com/science/article/pii/S0926985116306929}

\bibitem{Yan_IJNAMG_2019}
C.~Yan, Y.-Y. Jiao, H.~Zheng,
  \href{https://onlinelibrary.wiley.com/doi/abs/10.1002/nag.2937}{A
  three-dimensional heat transfer and thermal cracking model considering the
  effect of cracks on heat transfer}, International Journal for Numerical and
  Analytical Methods in Geomechanics 43~(10) (2019) 1825--1853.
\newblock \href {http://dx.doi.org/10.1002/nag.2937}
  {\path{doi:10.1002/nag.2937}}.
\newline\urlprefix\url{https://onlinelibrary.wiley.com/doi/abs/10.1002/nag.2937}

\bibitem{Bell_PYDEC_2012}
N.~Bell, A.~Hirani, Pydec: software and algorithms for discretization of
  exterior calculus, ACM Transactions on Mathematical Software 39~(1) (2012) 3.
\newblock \href {http://dx.doi.org/10.1145/2382585.2382588}
  {\path{doi:10.1145/2382585.2382588}}.

\bibitem{petsc-web-page}
S.~Balay, S.~Abhyankar, M.~F. Adams, J.~Brown, P.~Brune, K.~Buschelman,
  L.~Dalcin, A.~Dener, V.~Eijkhout, W.~D. Gropp, D.~Karpeyev, D.~Kaushik, M.~G.
  Knepley, D.~A. May, L.~C. McInnes, R.~T. Mills, T.~Munson, K.~Rupp, P.~Sanan,
  B.~F. Smith, S.~Zampini, H.~Zhang, H.~Zhang,
  \href{https://www.mcs.anl.gov/petsc}{{PETS}c {W}eb page},
  \url{https://www.mcs.anl.gov/petsc} (2019).
\newline\urlprefix\url{https://www.mcs.anl.gov/petsc}

\bibitem{hypre}
R.~D. Falgout, U.~M. Yang, hypre: A library of high performance
  preconditioners, in: P.~M.~A. Sloot, A.~G. Hoekstra, C.~J.~K. Tan, J.~J.
  Dongarra (Eds.), Computational Science --- ICCS 2002, Springer Berlin
  Heidelberg, Berlin, Heidelberg, 2002, pp. 632--641.

\bibitem{parasails}
E.~Chow, \href{https://doi.org/10.1137/S106482759833913X}{A priori sparsity
  patterns for parallel sparse approximate inverse preconditioners}, SIAM
  Journal on Scientific Computing 21~(5) (2000) 1804--1822.
\newblock \href
  {http://arxiv.org/abs/https://doi.org/10.1137/S106482759833913X}
  {\path{arXiv:https://doi.org/10.1137/S106482759833913X}}, \href
  {http://dx.doi.org/10.1137/S106482759833913X}
  {\path{doi:10.1137/S106482759833913X}}.
\newline\urlprefix\url{https://doi.org/10.1137/S106482759833913X}

\bibitem{Mamdouh_2018}
M.~S. Mohamed, A.~N. Hirani, R.~Samtaney,
  \href{https://doi.org/10.1080/15502287.2018.1446196}{Numerical convergence of
  discrete exterior calculus on arbitrary surface meshes}, International
  Journal for Computational Methods in Engineering Science and Mechanics 19~(3)
  (2018) 194--206.
\newblock \href
  {http://arxiv.org/abs/https://doi.org/10.1080/15502287.2018.1446196}
  {\path{arXiv:https://doi.org/10.1080/15502287.2018.1446196}}, \href
  {http://dx.doi.org/10.1080/15502287.2018.1446196}
  {\path{doi:10.1080/15502287.2018.1446196}}.
\newline\urlprefix\url{https://doi.org/10.1080/15502287.2018.1446196}

\end{thebibliography}


\begin{thebibliography}{0}
\bibitem{1}  S. Balay {\it et al.} PETSc Web page, https://www.mcs.anl.gov/petsc(2019).
\bibitem{2}  R. D. Falgout, U. M. Yang, hypre:  A library of high performance pre-conditioners,  in:  P.  M.  A.  Sloot,  A.  G.  Hoekstra,  C.  J.  K.  Tan,  J.  J.Dongarra (Eds.), Computational Science — ICCS 2002, Springer Berlin Heidelberg, Berlin, Heidelberg, 2002, pp. 632–641.
\bibitem{3}  E.   Chow,   A   priori   sparsity   patterns   for   parallel   sparse   approximate  inverse  preconditioners,  SIAM  Journal  on  Scientific  Computing  21  (5)  (2000)  1804–1822.
\bibitem{4} J.  R.  Shewchuk,  Triangle:  Engineering  a  2D  Quality  Mesh  Generator and Delaunay Triangulator, in:  M. C. Lin, D. Manocha (Eds.), Applied Computational Geometry: Towards Geometric Engineering, Vol. 1148 of Lecture Notes in Computer Science, Springer-Verlag, 1996, pp. 203–222, from the First ACM Workshop on Applied Computational Geometry.
\bibitem{5}  H. Si, Tetgen, a delaunay-based quality tetrahedral mesh generator, ACM Transactions on Mathematical Software 41 (2) (2015) 11. doi:10.1145/2629697.
\end{thebibliography}

\end{document}